\documentclass[a4paper]{article}

\usepackage{amsmath,graphicx}
\usepackage{xcolor}
\usepackage{multirow}
\usepackage{float}
\usepackage{cite}

\usepackage{booktabs,multirow}
\usepackage{adjustbox}
\renewcommand{\vec}[1]{\mathbf{#1}}
\renewcommand{\paragraph}[1]{\noindent\textbf{#1}}
\usepackage{url}
\usepackage{hyperref}
\usepackage{balance}

\usepackage{INTERSPEECH2021}

\title{SingAug: Data Augmentation for Singing Voice Synthesis with Cycle-consistent Training Strategy}
\name{Shuai Guo$^{1*}$\thanks{\scriptsize{$^*$Equal Contribution.}}, Jiatong Shi$^{2*}$, Tao Qian$^1$, Shinji Watanabe$^2$, Qin Jin$^{1 \dagger}$\thanks{\scriptsize{$^\dagger$Corresponding Author.}}
}
\address{
  $^1$School of Information, Renmin University of China, P.R.China \\
  $^2$Language Technologies Institute, Carnegie Mellon University, U.S.A. 
}
\email{\{shuaiguo, qiantao, qjin\}@ruc.edu.cn, jiatongs@cs.cmu.edu, shinjiw@ieee.org}

\begin{document}

\maketitle
\begin{abstract}
Deep learning based singing voice synthesis (SVS) systems have been demonstrated to flexibly generate singing with better qualities, compared to conventional statistical parametric based methods. However, neural systems are generally data-hungry and have difficulty to reach reasonable singing quality with limited public available training data. 
In this work, we explore different data augmentation methods to boost the training of SVS systems, including several strategies customized to SVS based on pitch augmentation and mix-up augmentation. 
To further stabilize the training, we introduce the cycle-consistent training strategy. Extensive experiments on two public singing databases demonstrate that our proposed augmentation methods and the stabilizing training strategy can significantly improve the performance on both objective and subjective evaluations.  
\end{abstract}
\noindent\textbf{Index Terms}: singing voice synthesis, data augmentation, cycle-consistent training strategy

\vspace{-5pt}

\section{Introduction}

\label{sec: intro}
In recent years, singing voice synthesis (SVS) has attracted much attention in both academic and industrial fields. The task takes music score and lyrics as input and generates natural singing voices. Over the last decade, deep neural network (DNN) based systems have achieved great performance in the synthesis field including SVS, and have shown their superiority over hidden Markov model (HMM) based models in both objective and subjective scores. In the beginning, the DNN model was proposed to predict the spectral information (e.g., Mel spectrogram, linear spectrogram, vocoder parameters) and started to outperform HMMs in mean opinion scores significantly \cite{nishimura2016singing, hono2018recent, tae2021mlp}. Later, variations of the neural networks, including recurrent neural networks (RNN) and convolutional neural networks (CNN), also demonstrated their power on acoustic modeling for singing voice \cite{blaauw2017neural, kim2018korean, nakamura2019singing, nakamura2020fast}. Architectures like generative adversarial network, were also shown to get reasonable performance by introducing various discriminators \cite{hono2019singing, chandna2019wgansing, liu2019score, choi2020korean, chen2020hifisinger, wu2020adversarially, lee2019adversarially, zhang2021visinger}.

As sequence-to-sequence models have become the dominant architectures in neural-based TTS, recent SVS systems have also adopted the encoder-decoder methods and achieved improved performance over the simple network structure (e.g., DNN, CNN, RNN) \cite{chen2020hifisinger, blaauw2020sequence, zhang2020durian, wu2020peking, gu2020bytesing, lu2020xiaoicesing, ren2020deepsinger, shi2021sequencetosequence}. In these methods, the encoders and decoders include Long-Short-Term Memory units (LSTM) with attention mechanism, multi-head self-attention (MHSA), and conformer blocks. However, in most cases, SVS has much less training data than TTS due to its high data annotation costs and more strict copyright requirements in the music domain.


We have observed three types of methods to mitigate the data scarcity problem, including: incorporating low-quality data, transfer learning, and regularization. Ren et al.~\cite{ren2020deepsinger} apply the data crawled from the Internet and multi-step preprocessing to train the end-to-end synthesis system. 
As singing has several similarities with speech, previous researches also investigate to apply transfer learning from speech \cite{valle2020mellotron, zhang2019learning}. 
However, the method requires parallel speech \& singing corpora, which is difficult to collect. 
Since limited data would more likely lead to over-fitting, a specific regularization loss based on perceptual entropy is proposed in \cite{shi2021sequencetosequence}, which improves the quality of synthesized singing. 
Since the perceptual entropy loss only acts as a regularization term, it has no decisive impact on the model training process, and there is still much room for improvements. 

Besides the methods mentioned above, data augmentation has been a simple and effective approach in other speech or singing related tasks \cite{hsu2019disentangling, hwang2020tts, huybrechts2020low, hwang2020mel, meng2021mixspeech, zhang2021pdaugment}. Following their insights, in this paper, we propose two simple data augmentation techniques, pitch augmentation and mix-up method, to improve the singing quality with limited training data. Moreover, a cycle-consistent predictor is introduced as an additional module to stabilize the training process with the augmentation techniques. Both our proposed data augmentation policies and the predictor module have demonstrated their superiority in both objective and subjective tests on two public singing datasets. Our model boosts the synthetic quality significantly, with the MOS performance improvement of 0.59 and 0.37 on Ofuton \cite{futon2021ofuton} and Opencpop \cite{wang2022opencpop}, respectively. \footnote{\scriptsize {The source code is publicly available at \url{https://github.com/SJTMusicTeam/Muskits} \cite{shi2022muskits}}}

\vspace{-5pt}

\begin{figure*}[t]
  \centering
  \includegraphics[width=0.9\linewidth]{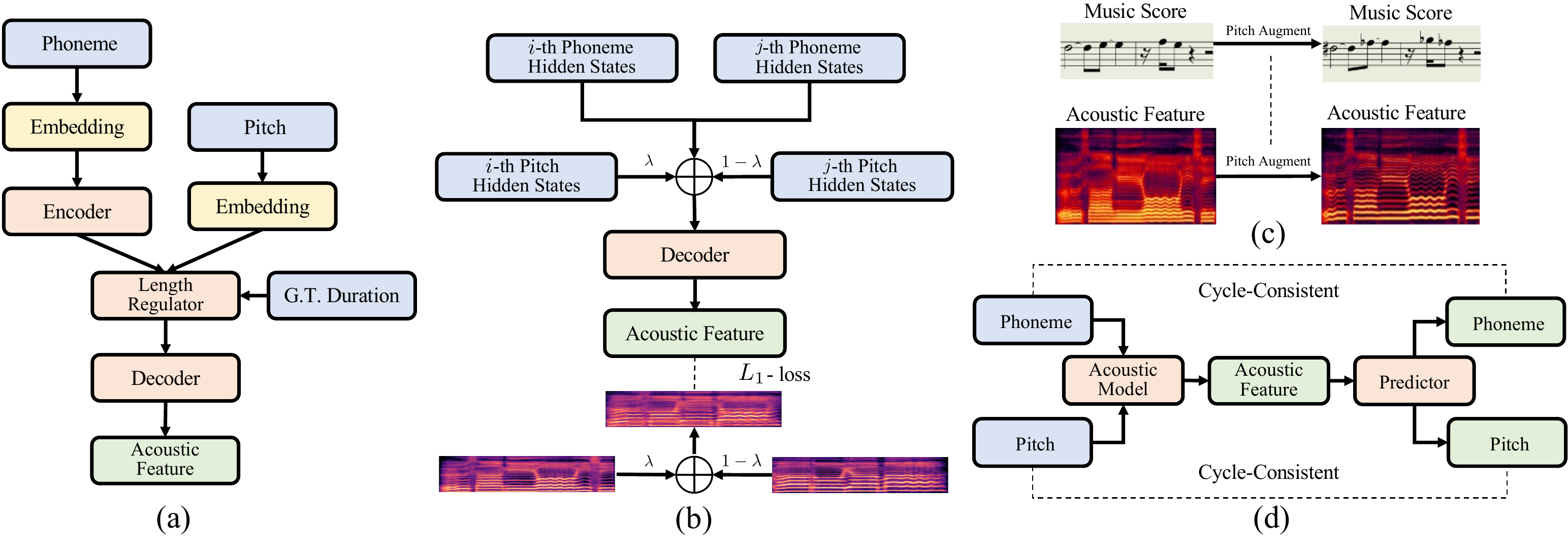}
    \vspace{-15pt}
  \caption{An overview of data augmentation and cycle-consistent training strategy for Singing Voice Synthesis. The blue blocks are the inputs, yellow and orange blocks are the main components of SVS model, green blocks are the outputs. (a) is the conventional pipeline of singing voice synthesis. (b) is the detailed process for mix-up augmentation. $\lambda$ is the combination weight in Eq.~(\ref{eq:mixup1}) and (\ref{eq:mixup3}). (c) is the diagram for pitch augmentation. The input music score and target acoustic feature need to be augmented correspondingly at the same time. (d) is the architecture for the cycle-consistent training strategy with predictor module.}

  \label{fig:framework}
\vspace{-15pt}
\end{figure*}

\section{Method}
\vspace{-2pt}

We denote the acoustic feature of a singing phrase as $Y \in \mathbb{R}^{T \times D^{a}}$, where $T$ is the number of frames in the sequence and $D^{a}$ is the acoustic feature dimension, and the music score $X := (X^{\text{ph}}, X^{\text{pi}}, X^{\text{beats}}  )$ including phoneme, pitch, and duration sequences are denoted as $X^{\text{ph}} \in \mathbb{N}^{T'}, X^{\text{pi}} \in \mathbb{N}^{T'}, X^{\text{beats}} \in \mathbb{N}^{T'}$, where $T'$ is the number of tokens.
Training the SVS model is to minimize the following loss:
\vspace{-5pt}
\begin{equation}
\begin{gathered}
    \mathcal{L}_{\text{svs}}^{\text{ori}}={L_1\left({f(X;\theta_{xy})},Y\right)}
\end{gathered}
\label{eq:svs_loss}
\end{equation}
where ${f(X;\theta_{xy})}$ denotes the acoustic model, which generates the prediction of acoustic features $\hat{Y}$ based on the music score (i.e., $(X^{\text{ph}}, X^{\text{pi}}, X^{\text{beats}})$). $\theta_{xy}$ refers to the trainable parameters of the acoustic model. $L_1(\cdot)$ denotes the L1 loss function. 

\vspace{-5pt}

\subsection{Non-Autoregressive SVS Framework}
\vspace{-4pt}
\label{sec: formulation}
The base framework of our SVS system (similar to \cite{lu2020xiaoicesing, wang2022opencpop}) is illustrated in Fig.~\ref{fig:framework} (a). The phoneme sequence $X^{\text{ph}}$ is converted into expanded phoneme hidden states $H^{\text{ph}} \in \mathbb{R}^{T \times D_h}$ after an embedding layer, an encoder, and a length regulator, where $D_h$ is the dimension of the hidden states $H^{\text{ph}}$. Meanwhile, $X^{\text{pi}}$ is passed to the pitch embedding layer and expanded by the same length regulator, resulting in expanded pitch hidden states $H^{\text{pi}} \in \mathbb{R}^{T \times D_h}$. In this work, similar to \cite{yi2019singing, wu2020adversarially, shi2021sequencetosequence}, we directly use the ground truth (G.T.) duration in our length regulator to 
avoid the duration modeling impact on our experiments. As the G.T. duration already contains the beats information, there is no need to use $X^{\text{beats}}$ in our SVS model. Based on the input of summing $H^{\text{ph}}$, $H^{\text{pi}}$, and the positional embedding, the decoder generates the prediction of acoustic feature $\hat{Y}$, from which the loss is calculated as in Eq.~\eqref{eq:svs_loss}.



Based on this framework, in the following subsections, we propose two singing augmentation methods, pitch and mix-up augmentation, to compensate the data scarcity issue for SVS. We further introduce a new cycle-consistent training strategy to improve the synthetic quality. 

\vspace{-5pt}

\begin{table*}
\centering
\caption{\label{tab: overview-results} Objective and subjective evaluation of SVS models with different settings on Ofuton and Opencpop databases. Augmentation policies include pitch augmentation (PA) and mix-up augmentation (MA). CC denotes as the acoustic model that is jointly trained with cycle-consistent predictor module. 
Both objective and subjective metrics are introduced in Sec.~\ref{ssec: exp-setup}.
}
\vspace{-10pt}
\begin{tabular}{l|l|ccccc|c}
\toprule
\textbf{Dataset}          & \multicolumn{1}{c|}{\textbf{Method}} & \textbf{MCD}$\downarrow$ & \textbf{LF0-RMSE}$\downarrow$ & \textbf{F0-CORR}$\uparrow$ & \textbf{ST\_ACC}$\uparrow$ & \textbf{VUV\_ERR}$\downarrow$ & \textbf{MOS}$\uparrow$     \\ \midrule
\multirow{6}{*}{Ofuton}   & Baseline                             & 6.88             & 0.103                 & 0.75             & 69.74             & 2.24                   & 2.65 ± 0.08          \\
                          & \quad + PA                                  & 6.55             & 0.106                 & 0.86             & 71.09                      & 2.38          & 2.87 ± 0.09          \\
                          & \quad + MA                                  & 6.65             & 0.101                 & 0.75             & 70.29                      & \textbf{2.07}                    & 2.89 ± 0.09          \\
                          & \quad + PA\&MA                              & 6.55             & 0.107                 & 0.87    & 71.74                      & 2.32                   & 2.98 ± 0.09          \\
                          & \quad + PA\&MA\&CC                       & \textbf{6.44}    & \textbf{0.100}        & \textbf{0.91}    & \textbf{71.88}             & 2.45          & \textbf{3.24 ± 0.09} \\ \cmidrule{2-8} 
                          & G.T.                                 & -                & -                     & -                & -                          & -                       & 4.60 ± 0.07          \\ \midrule
\multirow{6}{*}{Opencpop} & Baseline                             & 8.15             & 0.226                 & 0.87             & \textbf{74.46}             & 6.04                   & 2.89 ± 0.11          \\
                          & \quad + PA                                  & 7.85             & 0.219                 & 0.87             & 74.37                      & \textbf{5.68}          & 3.06 ± 0.11          \\
                          & \quad + MA                                  & 8.11             & 0.241                 & 0.88             & 74.41                      & 6.43                   & 2.92 ± 0.10          \\
                          & \quad + PA\&MA                              & 8.23             & 0.265                 & \textbf{0.89}    & 73.40                       & 6.59                   & 2.43 ± 0.09          \\
                          & \quad + PA\&MA\&CC                       & \textbf{7.76}    & \textbf{0.214}        & 0.86             & 74.30                       & 5.71                   & \textbf{3.26 ± 0.11} \\ \cmidrule{2-8} 
                          & G.T.                                 & -                & -                     & -                & -                          & -                       & 4.60 ± 0.09          \\
\bottomrule
\end{tabular}

\vspace{-15pt}
\end{table*}

\subsection{Singing Augmentation}
\vspace{-4pt}
\label{sec: augment}

\noindent \textbf{\textit{Mix-up Augmentation (MA)}}
Mix-up based data augmentation methods have been applied in supervised classification tasks \cite{zhang2018mixup, cheng2020advaug, meng2021mixspeech}. However, there are no related attempts in the regression tasks like TTS or SVS yet. 

The training pipeline of SVS with MA is shown in Fig.~\ref{fig:framework}~(b). Inspired by the MA in machine translation tasks \cite{cheng2020advaug}, we also adopt the MA in the embedding spaces to combine the music score information $X$. However, our MA methods are different from the MA in machine translation, as we perform the MA on the expanded hidden states $H^{\text{ph}}$ and $H^{\text{pi}}$ instead of on the input $X^{\text{ph}}$ and $X^{\text{pi}}$. The primary reason is that the samples selected for MA might have different duration for each note and leads to different expansions during length regulation. 
Specifically, on the basis of original training process, when applying MA, two samples $i$ and $j$ from a minibatch are selected randomly to form a mixture loss $\mathcal{L}_{\text{mix}}$. We denote $H \in \mathbb{R}^{T \times D_h}$ as the sum of $H^{\text{ph}}, H^{\text{pi}}$, and positional information. Then, the hidden states from the two samples are $H_i \in \mathbb{R}^{T_i \times D_h}$ and $H_j \in \mathbb{R}^{T_j \times D_h}$. We firstly pad hidden states on the right to make them in the same length $T_{\text{max}}=\mathrm{max}(T_i,T_j)$ and then interpolate them into $H_{\text{mix}} \in \mathbb{R}^{T_{\text{max}} \times D_h}$ on the left as:
\begin{equation}
\begin{gathered}
    H_{\text{mix}} = \lambda * H_{i} + (1-\lambda) * H_{j} ,
\end{gathered}
\label{eq:mixup1}
\end{equation}
where the weight $\lambda$ is sampled from a Beta distribution with hyper-parameter $\alpha$ (i.e., $\lambda \sim \mathrm{Beta}(\alpha, \alpha)$).

Given $H_{\text{mix}}$, the decoder predicts the acoustic feature as $\hat{Y}_{\text{mix}} \in  \mathbb{R}^{T_{\text{max}} \times D_a}$. $\hat{Y}_{\text{mix}}$ is used to compute the loss against the ground truth features $Y_i \in \mathbb{R}^{T_i \times D_a}$ and $Y_j \in \mathbb{R}^{T_j \times D_a}$.
%
%
With the same weight $\lambda$ used in Eq.~\eqref{eq:mixup1}, we define the mixture loss $\mathcal{L}_{\text{mix}}$ as:
\begin{equation}
\begin{gathered}
    \mathcal{L}_{\text{mix}} = \lambda * L_1(\hat{Y}_{\text{mix}},Y_{i}) + (1-\lambda) * L_1(\hat{Y}_\text{mix},Y_{j}) ,
\end{gathered}
\label{eq:mixup3}
\end{equation}
Finally, we utilize $w_{\text{mix}}$ to combine the original SVS loss $\mathcal{L}_{\text{svs}}^{\text{ori}}$ and the additional mix-up loss $\mathcal{L}_{\text{mix}}$:
\begin{equation}
\begin{gathered}
    \mathcal{L}_{\text{svs}} = (1-w_{\text{mix}}) * \mathcal{L}_{\text{svs}}^{\text{ori}} + w_{\text{mix}} * \mathcal{L}_{\text{mix}} .
\end{gathered}
\label{eq:mixup4}
\end{equation}




\noindent \textbf{\textit{Pitch Augmentation (PA)}}
According to the chromatic scale \cite{quinn2019tonal}, the musical pitches can be scaled to 12 unique semitones. They are designed to be equally-spaced in twelve-tone equal temperament, which is one of the dominant temperaments \cite{benward2014music}. Each semitone differs by $\sqrt[12]{2}$ in frequency domain (Hz), equals to 1.059 approximately.

For pitch augmentation, we employ WORLD vocoder \cite{morise2016world} to get the corresponding singing voice after semitone adjustments as shown in Fig.~\ref{fig:framework}(c). The input information of the WORLD vocoder consists of the fundamental frequency F0, the spectral envelope SP (harmonic spectral envelope) and the aperiodic signal AP (aperiodic spectral envelope). By multiplying or dividing $\sqrt[12]{2}$ to F0 sequence and keeping the SP and AP unchanged, we can easily obtain the time domain singing waveform signal corresponding to its semitone adjustments. 
Different from pitch augmentation or pitch shifting discussed in \cite{blaauw2017neural, zhang2021pdaugment, zhang2022wesinger}, in our strategy, the entire singing phrase is processed with a fixed amount of semitone over raw audios, and the music score is processed accordingly. Based on the chromatic scale, the melody can be kept as the same over another tonality. 

\vspace{-5pt}

\subsection{Cycle-consistent training strategy (CC)} 
\vspace{-4pt}
\label{ssec: SVS Framework with Predictor}




As discussed in Section~\ref{sec: augment}, our proposed augmentation policies would significantly increase the diversity of training samples.
However, both MA and PA apply modification directly to acoustic features, which may introduce artifacts and noises to the original signal. For PA, the noise may come from the WORLD vocoder when modifying target singing signals, as its quality still has a gap with the real singing voice. For MA, $i$-th and $j$-th samples can be regarded as the noise of each other. These artifacts and noises may disrupt the training of the SVS acoustic model.
Therefore, we propose our new cycle-consistent training strategy which is shown in Fig.~\ref{fig:framework}~(d). It applies an additional predictor to stabilize the network and makes it resistant to the potential noise from augmentation. 

Under the cycle-consistent training strategy with predictor module, the score information loss $\mathcal{L}_{\text{si}}$ helps the predictor module train steadily from the G.T. acoustic feature and music score:
\vspace{-3pt}
\begin{equation}
\begin{gathered}
    \mathcal{L}_{\text{si}}={\mathrm{CrossEntropy}\left({g(Y;\theta}_{yx}),X\right)} ,
\end{gathered}
\label{eq:scoreInfo_loss}
\end{equation}
where ${g(Y;\theta}_{yx})$ denotes the predictor module, which takes the G.T. acoustic feature as input, produces the prediction of expanded phoneme and pitch sequences. $\theta_{yx}$ represents trainable parameters of the predictor module. Given any paired $(x,y)$, we use the Cross-Entropy loss to train the predictor module.

Similar to models for lyric recognition and music transcription, the predictor takes the predicted acoustic feature as input and predicts their phoneme and pitch sequences. The related formulation is defined as follows, which is the composition function of Eq.~(\ref{eq:svs_loss}) and Eq.~(\ref{eq:scoreInfo_loss}):
\vspace{-3pt}
\begin{equation}
\begin{gathered}
    \mathcal{L}_{\text{pd}}={\mathrm{CrossEntropy}\left({g(f(X;\theta_{xy});\theta}_{yx}),X\right)} ,
\end{gathered}
\label{eq:predictor_loss}
\end{equation}
when calculating the loss $\mathcal{L}_{\text{pd}}$ of the predictor module, it takes the predicted acoustic feature as input, produces the prediction of phoneme and pitch sequences. It aims to help the acoustic model keep the cycle consistency of music score information. 

At last, we combine all the losses in Eq.~(\ref{eq:svs_loss}), ~(\ref{eq:scoreInfo_loss}) and ~(\ref{eq:predictor_loss}) using different weights, including $w_{\text{svs}}$, $w_{\text{si}}$ and $w_{\text{pd}}$. The combined loss $\mathcal{L}$ is used for our cycle-consistent SVS training:
\begin{equation}
\begin{gathered}
    \mathcal{L} = w_{\text{svs}} * \mathcal{L}_{\text{svs}} + w_{\text{si}} * \mathcal{L}_{\text{si}} + w_{\text{pd}} * \mathcal{L}_{\text{pd}} .
\end{gathered}
\label{eq:framework_loss}
\end{equation}
\vspace{-21pt}

\section{Experiments}
\vspace{-4pt}
\subsection{Datasets}
To evaluate the effectiveness of our methods, we conduct related experiments on both Ofuton \cite{futon2021ofuton} and Opencpop \cite{wang2022opencpop} databases. Ofuton dataset is a public Japanese male singing voice corpora, which has 56 Japanese songs (61 minutes) in total. 
Since Ofuton corpora has no official segmentation, we split each song into several singing phrases, resulting in 547 phrases for training, 58 for validation, and 70 for testing. The splitting is based on the silence between lyrics. Opencpop dataset is a public Mandarin female singing voice corpus. It has 100 popular Mandarin songs (5.2 hours), which are performed by a female singer. 
As for segmentation of Opencpop dataset, we follow the official split. For preprocessing, we down-sample the songs to a sampling rate of 24k Hz and extract the 80-dim Mel spectrogram as the target acoustic features in our SVS system. The Mel spectrogram is extracted with 12.5 ms frame-shift and 50 ms frame-length.

\vspace{-5pt}

\subsection{Experiment Setups}
\vspace{-4pt}
\label{ssec: exp-setup}
For the SVS acoustic model in our paper, we follow the architecture of Fastspeech \cite{ren2019fastspeech}, which is a encoder-decoder based sequence-to-sequence (Seq2Seq) model widely used in both TTS and SVS domain. The encoder has 6 blocks of encoder layers. Each encoder layer consists of a 384-dimension, four-heads self-attention layer and 1D-convolution feed-forward module. The convolutional layer in the feed-forward module consists of 1536 filters with shape 1 × 1. The embedding size of phoneme and pitch are 384. The decoder has 6 blocks of decoder layers, which keep the same model structure as encoder layer. LayerNorm and dropout are adopted in both encoder and decoder layer. The dropout rate is set as 0.1. The post-net module is utilized after decoder to refine the output Mel spectrogram by predicting the residual. The detailed model structure of post-net module follows the Tacotron 2 \cite{shen2018natural}.

For the predictor module, 4 blocks of predictor layers are included in both phoneme predictor module and pitch predictor module. Each predictor layer consists of a 80-dimension, four-heads self-attention layer and 1D-convolution feed-forward module. The convolutional layer in the feed-forward module consists of 512 filters with shape 1 × 1. The dropout rate is set as 0.1. After the predictor layer, a linear mapping is adopted to produce the prediction of phoneme or pitch sequence.

For the vocoder, we utilize the HiFi-GAN vocoder \cite{kong2020hifi} to generate the singing waveform from predicted Mel spectrogram. The vocoder is pre-trained with the G.T. audio and Mel spectrogram pairs for 300k steps in advance.

In the training stage, the Adam optimizer with 0.001 learning rate and noam warm-up policy \cite{vaswani2017attention} are utilized. Global mean normalization is applied for acoustic features. We train the acoustic models for 500 epochs in Ofuton dataset and 250 epochs in Opencpop dataset. The models with the lowest validation loss $\mathcal{L}_{\text{svs}}$ are chosen for testing. During inference, we use the G.T. duration information to expand the encoder outputs.
For PA, we randomly shift semitones in a phrase by \{-1, 0, 1\}. For MA, the proportion of mix-up samples within a batch is set as 0.15. The weight of mix-up loss $w_{\text{mix}}$ is 0.1 in Ofuton database and Opencpoop database. The $\alpha$ of Beta distribution which controls the combination weight $\lambda$ in Eq.~(\ref{eq:mixup1}) is set as 0.5. 
In Ofuton database, the CC's weights to combine the loss terms mentioned in Equation (2) are set as $w_{\text{svs}}=0.7$, $w_{\text{si}}=0.2$, $w_{\text{pd}}=0.1$. While in the Opencpop database, the weights are set as 0.85, 0.1, 0.05 respectively.

For the objective evaluation, we utilize five metrics, including Mel-cepstrum distortion (MCD), log-F0 root mean square error (LF0\_RMSE), Pearson correlation coefficients of F0 measures (F0\_CORR), semitone accuracy (ST\_ACC) and voice/unvoiced error rate (VUV\_ERR). We use ST\_ACC to allow some tolerance over F0 differences, based on music theory and human hearing perception \cite{quinn2019tonal}.

For the subjective evaluation, we conduct the Mean opinion score (MOS) test to verify the effectiveness of our methods. We invite 25 listeners ranging from musicians and non-professionals. Listeners are asked to give their opinion score from one (non-intelligible) to five (excellent naturalness) in a blind and order-randomized fashion. For each setting, 15 samples are chosen randomly from the test set.


\vspace{-5pt}

\begin{table}[]
\centering
\caption{\label{tab: pitch ablation} The comparison of different ways to choose the PA shifting factor on Opencpop dataset.}
\vspace{-10pt}
\begin{tabular}{c|ccc}
\toprule
\textbf{Method} & \textbf{MCD$\downarrow$} & \textbf{ST\_ACC$\uparrow$} & \textbf{VUV\_ERR$\downarrow$} \\
\midrule
P-adaptive & 8.26             & 73.29                     & 6.77                   \\
P1         & \textbf{7.85}             & \textbf{74.37}                     & \textbf{5.68}                   \\
P2         & 8.08             & 73.92                     & 6.06    \\             
\bottomrule
\end{tabular}
\vspace{-10pt}
\end{table}

\begin{table}[]
\centering
\caption{\label{tab: mixup ablation} The comparison of different weights in MA on Opencpop dataset.}
\vspace{-10pt}
\begin{tabular}{c|ccc}
\toprule
\textbf{Method} & \textbf{MCD$\downarrow$} & \textbf{ST\_ACC$\uparrow$} & \textbf{VUV\_ERR$\downarrow$} \\
\midrule
$w_{mix}=0.1$    & \textbf{8.11}             & \textbf{74.41}            & 6.43                   \\
$w_{mix}=0.2$    & 8.15             & 73.60                      & 6.43         \\       
$w_{mix}=0.3$    & 8.20             & 74.25                      & \textbf{6.36}         \\ 
\bottomrule
\end{tabular}
\vspace{-16pt}
\end{table}

\subsection{Comparison with the Baseline}
\vspace{-4pt}

Table~\ref{tab: overview-results} presents the results on Ofuton and Opencpoop databases using our proposed augmentation methods and cycle-consistent training strategy. The detailed structure of the baseline model is described in Section~\ref{ssec: exp-setup}. 

On the Ofuton database, we can observe a general trend that PA, MA, PA\&MA\&CC improve the performance on both subjective and objective scores, expect on the VUV\_ERR metric. 
This may be due to the artifacts introduced from the semitone shift on the training data when PA is applied.

On the Opencpop database, although  PA, MA, or PA\&MA\&CC achieves the best scores in subjective and objective evaluations, we do observe more results fluctuations in F0-based measures. It may indicate that the augmentation methods could sacrifice limited performance in F0 accuracy of the music score to reach a significant gain on naturalness of the signal. 

Compared to results with single augmentation method (PA or MA), combining both augmentations (PA\&MA) can enhance the performance on Ofuton, but not on Opencpop. However, after applying CC, the system reaches the best MOS, which aligns with our assumption that the introduction of CC can improve the stability over potential noises from augmentation policies. 

\vspace{-5pt}

\subsection{Ablation Studies}
\vspace{-4pt}
We conduct experiments to investigate the impact of different hyper-parameters on the model. All of the ablation studies are conducted on the Opencpop dataset.

\noindent \textbf{A. Pitch shifting factor in PA} Table~\ref{tab: pitch ablation} presents the results with different PA shifting. P-adaptive means that we make the average pitch of samples closer to the mean of Opencpop dataset, e.g. if the mean pitch of the sample and dataset are 63 and 65, the shifting will range in \{0, 1, 2\}. P1 is the default setting we used in Sec.~\ref{ssec: exp-setup}.  The approach of P2 is similar to P1, while the shifting ranges in \{-2, -1, 0, 1, 2\}. Among these three methods of pitch augmentation, P1 achieves the best performance on all objective metrics. As shown in Table~\ref{tab: pitch ablation}, the variation range of PA is not proportional to the synthetic quality. Too much variation in PA may reduce the effectiveness of the acoustic model.

\begin{table}[]
\centering
\caption{\label{tab: cycle-PA-MA ablation} The comparison of different weights in cycle consistent training strategy on Opencpop dataset. CC1 denotes the setting of $w_{\text{svs}}$=0.7, $w_{\text{si}}$=0.2, $w_{\text{pd}}$=0.1. CC2 denotes the setting of $w_{\text{svs}}$=0.85, $w_{\text{si}}$=0.1, $w_{\text{pd}}$=0.05. CC3 denotes the setting of $w_{\text{svs}}$=1, $w_{\text{si}}$=1, $w_{\text{pd}}$=1.}
\vspace{-10pt}
\begin{tabular}{c|ccc}
\toprule
\textbf{Method} & \textbf{MCD$\downarrow$} & \textbf{ST\_ACC$\uparrow$} & \textbf{VUV\_ERR$\downarrow$} \\
\midrule
CC1 + PA + MA & 8.25             & 73.89                     & 6.62                   \\
CC2 + PA + MA & \textbf{7.76}    & \textbf{74.30}                      & \textbf{5.71}          \\
CC3 + PA + MA & 8.71             & 73.83                     & 6.45         \\             
\bottomrule
\end{tabular}
\vspace{-10pt}
\end{table}

\begin{table}[]
\centering
\caption{\label{tab: cycle with PA or MA ablation} The ablation study of cycle consistent SVS framework with pitch or mix-up augmentation on Opencpop dataset.}
\vspace{-10pt}
\begin{tabular}{c|ccc}
\toprule
\textbf{Method} & \textbf{MCD$\downarrow$} & \textbf{ST\_ACC$\uparrow$} & \textbf{VUV\_ERR$\downarrow$} \\
\midrule
CC2         & 8.06             & 73.67                     & 6.63                   \\
CC2 + PA          & 8.11             & 73.46                     & 6.44                   \\
CC2 + MA    & 8.04             & 74.28            & 6.20           \\       
CC2 + PA + MA    & \textbf{7.76}             & \textbf{74.30}            & \textbf{5.71}           \\    
\bottomrule
\end{tabular}
\vspace{-16pt}
\end{table}

\noindent \textbf{B. Mixing weights in MA} Table~\ref{tab: mixup ablation} presents the results with different weights applied in MA. We change the weights of mix-up augmentation $w_{mix}$ in Eq.~(\ref{eq:mixup4}). When $w_{mix}=0.1$, the model reaches the optimal value on MCD and ST\_ACC. The model with $w_{mix}=0.3$ achieves the best VUV\_ERR performance. Note that the trend in MA is similar to that in PA, large weights of $w_{mix}$ may damage the synthetic performance.

\noindent \textbf{C. Different weights in CC} Table~\ref{tab: cycle-PA-MA ablation} presents the results with different weights in CC with predictor. We verify the models with the combination of PA and MA (both with default settings). Among different weights in Table~\ref{tab: cycle-PA-MA ablation}, CC2 ( $w_{\text{svs}}=0.85$, $w_{\text{
si}}=0.1$, $w_{\text{pe}}=0.05$) achieves the best performance on all objective metrics. Compared to the methods using PA and MA in Table~\ref{tab: overview-results} on Opencpop dataset, combining the CC predictor module does not necessarily enhance the synthetic quality of the acoustic model. It indicates that the weights of CC predictor module might need some tuning for usage.

\noindent \textbf{D. Combination of CC with PA or MA} Table~\ref{tab: cycle with PA or MA ablation} presents the results of the combination of CC with PA or MA, among which CC+PA+MA achieves the best performance on all objective metrics. Compared to the baseline model in Table~\ref{tab: overview-results}, the methods with CC improve the MCD metric regardless of whether PA or MA is used. Moreover, combing the CC with PA and MA could bring further improvements on synthetic quality, which demonstrates the effectiveness of our methods. 
\vspace{-5pt}

\section{Conclusions}
\vspace{-4pt}
In this paper, we propose two data augmentation methods to mitigate the data scarcity issue for SVS. We further introduce a new training strategy that jointly trains the SVS network with a cycle-consistent predictor module for music and lyrics information. Our proposed methods bring improvements on both subjective and objective metrics on two different datasets, e.g. gaining 0.59 and 0.37 MOS improvement on Ofuton \cite{futon2021ofuton} and Opencpop \cite{wang2022opencpop}, respectively. We will explore extending the proposed framework into unsupervised fashion in the future work, e.g., using dual learning or self-supervised learning.

\vspace{-4pt}
\section{Acknowledgement}
\vspace{-4pt}
This work was supported by the National Natural Science Foundation of China (No. 62072462) and the National Key R\&D Program of China (No. 2020AAA0108600).

\vfill\pagebreak
\balance

\bibliographystyle{IEEEtran}

\bibliography{mybib}

\end{document}